\documentstyle[12pt]{article}
\makeatletter
\begin{document}
%\begin{titlepage}
%\begin{flushright}
%math.QA/9907134
%\end{flushright}
%\vskip.3in

\title{
{ \huge   Modular transformation and 
twist between trigonometric limits of $sl(n)$
elliptic R-matrix}}

\author{\bf \large \\
Wen-Li Yang$^{a,b}$\thanks{E-mail address: wlyang@phy.nwu.edu.cn}~~ 
and ~~Yi Zhen$^{a}$\thanks{E-mail address: zheny@phy.nwu.edu.cn} \\
{\small $~^{a}$Institute of Modern Physics, Northwest University, 
Xian 710069, China }\\
{\small $~^{b}$ Physikalishes Institut der Universit\"at Bonn, Nussallee 12, 
 53115 Bonn, Germany}
}
\maketitle

\begin{abstract}

We study the modular transformation of ${\bf Z}_n$-symmetric
elliptic R-matrix and  construct the twist between the trigonometric
degeneracy of the  elliptic R-matrix.

\vskip 1cm
\noindent{\bf Mathematics Subject Classifications (1991):} 
    17B37, 81R10, 81R50, 16W30.

\end{abstract}

%  Greek letters

\def\a{\alpha}
\def\b{\beta}
\def\d{\delta}
\def\e{\epsilon}
\def\ve{\varepsilon}
\def\g{\gamma}
\def\k{\kappa}
\def\l{\lambda}
\def\o{\omega}
\def\t{\theta}
\def\s{\sigma}
\def\D{\Delta}
\def\L{\Lambda}

\def\R{\overline{R}}
\def\S{\overline{S}}
\def\Sl{{sl(n)}}
\def\DR{{R_{\xi}^{DY}}}
\def\VR{{R^{Y}}} 
\def\hS{{\widehat{sl(n)}}}
\def\hG{{\widehat{gl(N|N)}}}
\def\R{{\cal R}}
\def\hR{{\hat{\cal R}}}
\def\C{{\cal C}}
\def\P{{\bf P}}
\def\Z2{{{\bf Z}_2}}
\def\Z{{\bf Z}_n}
\def\T{{\cal T}}
\def\H{{\cal H}}
\def\F{{\cal F}}
\def\V{\overline{V}}
\def\trho{{\tilde{\rho}}}
\def\tphi{{\tilde{\phi}}}
\def\tT{{\tilde{\cal T}}}
\def\uqsnh{{U_q[\widehat{sl(N|N)}]}}
\def\uqgnh{{U_q[\widehat{gl(N|N)}]}}
\def\uq1h{{U_q[\widehat{gl(1|1)}]}}
\def\uqg2h{{U_q[\widehat{gl(2|2)}]}}

% Shorthands for \begin{equation} and the like

\def\beq{\begin{equation}}
\def\eeq{\end{equation}}
\def\bea{\begin{eqnarray}}
\def\eea{\end{eqnarray}}
\def\ba{\begin{array}}
\def\ea{\end{array}}
\def\no{\nonumber}
\def\lt{\left}
\def\rt{\right}
\newcommand{\bq}{\begin{quote}}
\newcommand{\eq}{\end{quote}}

\newtheorem{Theorem}{Theorem}
\newtheorem{Definition}{Definition}
\newtheorem{Proposition}{Proposition}
\newtheorem{Lemma}{Lemma}
\newtheorem{Corollary}{Corollary}
\newcommand{\proof}[1]{{\bf Proof. }
        #1\begin{flushright}$\Box$\end{flushright}}

\newcommand{\sect}[1]{\setcounter{equation}{0}\section{#1}}
\renewcommand{\theequation}{\thesection.\arabic{equation}}

\sect{Introduction\label{intro}}
Universal twists connecting quantum (super) algebras to elliptic 
(super) algebras have been constructed in 
\cite{Bab96,Fro97,Jim97,Zha99,Gou98}. Recently, universal twists
connecting double Yangian $DY(sl(2))_c$ to deformed double
Yangian $DY_{\xi}(sl(2))_c$\cite{Arn99} (or
$A_{\hbar,\eta}(\widehat{sl_2})$ 
\cite{Jim96,Kho98}), has been given in \cite{Arn00}. They
show the quasi-Hopf structure\cite{Dri90} of elliptic (super) algebras
and deformed Yangian. 

If $({\cal A},\D, {\cal R})$ defines  a quasi-triangular Hopf
(super) algebra 
and the universal twist $\F \in {\cal A}\otimes {\cal A}$ satisfies the 
cocycle-like  relations ( For the details we will refer the reader to
\cite{Jim97,Zha99}), $({\cal A},\D^{\F}, {\cal R}^{\F})$ with
\bea
\D^{\F}(\cdot)=\F\D(\cdot)\F^{-1}, ~~~
{\cal R}^{\F}=F_{21}{\cal R}\F_{12}^{-1},
\eea
defines a quasi-triangular quasi-Hopf (super) algebra. As for the
R-matrices interpreted as evaluation representation of universal ones,  
it becomes 
\bea
R^{\F}=F_{21}~R~F_{12}^{-1},
\eea
where the particular matrix $F$ is the evaluation representation of 
$\F$.

It is well-known that there exist two kind trigonometric degeneracy limits
of elliptic R-matrix: scaling limit which is the R-matrix of deformed 
double Yangian $DY_{\xi}(sl(n))_c$\cite{Arn99} and ordinary limit which 
is also related to another type deformed Yangian (e.g
$DY_{\xi}^{V6}$\cite{Jim97}). In this letter, we will construct the twist 
between these two trigonometric R-matrices. In  special case $n=2$, our 
result coincides with the result of Arnaudon's\cite{Arn00}. 

\sect{Trigonometric  limits of elliptic R-matrix}

\subsection{The $\Sl$ elliptic R-matrix}

Let $n$ be an integers and $n\leq 2$, $w\in \C$ and $Imw> 0$, 
$\xi$ take real value and $\xi>0$, $\tau\in\C$ and $Im\tau >0$. 
Set ${\bf V}$ be a n-dimensional vector space with standard basis 
$\{e_j\}_{j\in\Z}$. Introduce  $n\times n$ matrices $h$, $g$ 
 and $I_{\a}$ with $\a\in \Z\otimes\Z$ by 
\bea
&&he_j=e_{j+1},~~~~ge_j=\o^je_j,~~~j\in\Z,\no\\
&&I_{\a}\equiv I_{\a_1,\a_2}=h^{\a_1}g^{\a_2},\no
\eea
where $\o=exp\{\frac{2i\pi}{n}\}$. 
Define the elliptic functions 
\bea
&&\t \left[
\begin{array}{l}
a\\b
\end{array}
\right](z,\tau)=\sum_{m\in Z}exp\{i\pi[(m+a)^2\tau+2(m+a)(z+b)]\},\no\\
&&\s_{\a}(z,\tau)=\s_{(\a_1,\a_2)}(z,\tau)=
\t \left[
\begin{array}{l}
\frac{1}{2}+\frac{\a_1}{n}\\
\frac{1}{2}+\frac{\a_2}{n}
\end{array}
\right](z,\tau).\no
\eea
The $\Z$-symmetric R-matrix can be defined as 
\bea
\S(z,w,\tau)=\frac{\s_0(w,\tau)}
{\s_0(z+w,\tau)}\sum_{\a}W_{\a}(z,\tau)I_{\a}\otimes
I_{\a}^{-1},\label{ZR}
\eea
where 
\bea
W_{\a}(z,\tau)=\frac
{\s_{\a}(z+\frac{w}{n},\tau)}
{n\s_{\a}(\frac{w}{n},\tau)}.\no
\eea
The elements of $\Z$-symmetric R-matrix can be expressed
explicitly\cite{Ric86}
\bea
\{\S(z,w,\tau)\}^{kl}_{ij}&=&\frac{\s_0(w,\tau)}{\s_0(z+w,\tau)}
~\frac{\displaystyle\prod_{j=0}^{n-1}
\t \left[
\begin{array}{c}
\frac{1}{2}+\frac{j}{n}\\
\frac{1}{2}
\end{array}
\right](z,n\tau)}
{\displaystyle\prod_{j=1}^{n-1}\t \left[
\begin{array}{c}
\frac{1}{2}+\frac{j}{n}\\
\frac{1}{2}
\end{array}
\right](0,n\tau)}\no\\
&&~~~~\times 
\frac{\t \left[
\begin{array}{c}
\frac{1}{2}+\frac{l-k}{n}\\
\frac{1}{2}
\end{array}
\right](z+w,n\tau)}
{\t \left[\begin{array}{c}
\frac{1}{2}+\frac{i-k}{n}\\
\frac{1}{2}
\end{array}
\right](w,n\tau)
\t \left[\begin{array}{c}
\frac{1}{2}+\frac{l-i}{n}\\
\frac{1}{2}
\end{array}
\right](z,n\tau)}.
\eea

Introduce  $\Sl$ elliptic R-matrix 
\bea 
S(z)\equiv S(z,w,\tau
)=x^{2\frac{z}{w}(\frac{1}{n}-1)}\frac{g_1(\frac{z}{w})}
{g_1(-\frac{z}{w})}\S(z
,w,\tau),\label{ER}
\eea
with  $x=e^{i\pi w}$ and 
\bea
g_1(v)=\frac{\{x^{2v}x^2\} \{x^{2v}x^{2n+2\xi-2}\}} 
{\{x^{2v}x^{2n}\} \{x^{2v}x^{2\xi}\}},
\eea
where $\{z\}=(z;e^{2i\pi\tau},x^{2n})$ and $(z;p_1,\cdots,p_m)\equiv 
\prod^{\infty}_{\{n_i\}=0}(1-zp_1^{n_1}\cdots p^{n_m}_m)$. 

The  $\Sl$ elliptic R-matrix $S(v)$ satisfies the following
properties\cite{Fan98}
\bea
{\rm Yang-Baxter~equation}~
&:&~~S_{12}(v_1-v_2)S_{13}(v_1-v_3)S_{23}(v_2-v_3)\no\\
&&~~~~~~=
S_{23}(v_2-v_3)S_{13}(v_1-v_3)S_{12}(v_1-v_2),\no\\
{\rm Unitarity}~&:&~~S_{12}(v)S_{21}(-v)=1,\no\\
{\rm Crossing-Unitarity}~&:&~~S_{12}(v)^{t_2}S_{21}(-v-n)^{t_2}=1,\no
\eea

\subsection{Scaling limit of the $\Sl$ elliptic R-matrix}
The scaling limit of the R-matrix (\ref{ER}) is taken by
\bea 
\frac{z}{\tau}=\frac{i\b}{\hbar\xi},~~
\frac{w}{\tau}=\frac{1}{\xi}, ~~w\longrightarrow 0+,
~~{\rm
with}~ \b,\xi~{\rm and}~\hbar~{\rm
fixed}.\label{scaling}
\eea
The above limit should be understood as $w$ go to $0$ with the $Imw$ from 
$0+$.
Noting the properties of elliptic functions under the scaling limit 
(\ref{scaling})
\bea
\frac{\t \left[\begin{array}{c}
\frac{1}{2}+a\\
\frac{1}{2}
\end{array}
\right](\frac{i\b w}{\hbar},n\xi w)}
{\t \left[\begin{array}{c}
\frac{1}{2}+b\\
\frac{1}{2}
\end{array}
\right](\frac{i\b w}{\hbar},n\xi w)}~
\stackrel{w\rightarrow 0+}{\longrightarrow}
~\frac{e^{i\pi(a-b)}sin\pi(\frac{i\b w}{n\hbar\xi}+a)}
{sin\pi(\frac{i\b w}{n\hbar\xi}+b)},\\
\frac{\t \left[\begin{array}{c}
\frac{1}{2}+a\\
\frac{1}{2}
\end{array}
\right](\frac{i\b w}{\hbar},\xi w)}
{\t \left[\begin{array}{c}
\frac{1}{2}+b\\
\frac{1}{2}
\end{array}
\right](\frac{i\b w}{\hbar},\xi w)}~
\stackrel{w\rightarrow 0+}{\longrightarrow}
~\frac{e^{i\pi(a-b)}sin\pi(\frac{i\b w}{\hbar\xi}+a)}
{sin\pi(\frac{i\b w}{\hbar\xi}+b)},
\eea
we can obtain that in the scaling limit the matrix elements of 
(\ref{ER}) become
\bea
\{\DR\}^{kl}_{ij}(\b,\xi;\hbar)&\stackrel{def}{=}&
\lim_{w\rightarrow 0+}
S^{kl}_{ij}(\frac{i\b w}{\hbar},w,\xi w)\label{slim}\\
&=&
\lim_{w\rightarrow 0+}x^{\frac{2i\b(1-n)}{n\hbar}}
\frac{g_1(\frac{i\b}{\hbar})}{g_1(-\frac{i\b}{\hbar})}
\times\lim_{w\rightarrow 0+}
\S^{kl}_{ij}(\frac{i\b w}{\hbar},w,\xi w)\no\\
&=&\k(\b)\{\prod_{\a=1}^{n-1}
\frac{sin\pi(\frac{i\b}{n\hbar\xi}+\frac{\a}{n})}
{sin\pi(\frac{\a}{n})}\}
\frac{sin\pi(\frac{1}{\xi})
sin\pi(\frac{i\b}{n\hbar\xi})sin\pi(\frac{i\b}{n\hbar\xi}+
\frac{1}{n\xi}+\frac{l-k}{n})}
{sin\pi(\frac{i\b}{\hbar\xi}+\frac{1}{\xi})
sin\pi(\frac{1}{n\xi}+\frac{i-k}{n})sin\pi(\frac{i\b}{n\hbar\xi}+
\frac{l-i}{n})},\no\\
&=&\k(\b)
\frac{sin\pi(\frac{1}{\xi})
sin\pi(\frac{i\b}{\hbar\xi})sin\pi(\frac{i\b}{n\hbar\xi}+
\frac{1}{n\xi}+\frac{l-k}{n})}
{n~sin\pi(\frac{i\b}{\hbar\xi}+\frac{1}{\xi})
sin\pi(\frac{1}{n\xi}+\frac{i-k}{n})sin\pi(\frac{i\b}{n\hbar\xi}+
\frac{l-i}{n})},\no
\eea
where the scalar factor $\k(\b)$ is given
\bea
\k(\b)=exp\{-2\int_0^{\infty}\frac{sh(n-1)\hbar t~sh(\xi-1)\hbar
t~sh2i\b t}{sh\hbar\xi t~shn\hbar t}\frac{dt}{t}\},\no
\eea
and $\k(\b)$ can be re-expressed in terms of double sine
function\cite{Arn99}.
In the deriving, we have used the identity
\bea 
\prod^{n-1}_{j=1}\frac{sin(x+\frac{j\pi}{n})}{sin(\frac{j\pi}{n})}
=\frac{sinnx}{n~sinx}.\no
\eea
In the particular case n=2, one recovers explicitly\cite{Kon97,Arn99}
\bea
\DR(\b,\xi;\hbar)=\k(\b)
\left(\begin{array}{cccc}
\frac{cos\frac{\pi}{2\xi}~cos\frac{i\pi\b}{2\hbar\xi}}
{cos\pi(\frac{i\b}{2\hbar\xi}+\frac{1}{2\xi})}
&0&0&
-\frac{sin\frac{\pi}{2\xi}~sin\frac{i\pi\b}{2\hbar\xi}}
{cos\pi(\frac{i\b}{2\hbar\xi}+\frac{1}{2\xi})}
\\
0&\frac{cos\frac{\pi}{2\xi}~sin\frac{i\pi\b}{2\hbar\xi}}
{sin\pi(\frac{i\b}{2\hbar\xi}+\frac{1}{2\xi})}
&\frac{sin\frac{\pi}{2\xi}~cos\frac{i\pi\b}{2\hbar\xi}}
{sin\pi(\frac{i\b}{2\hbar\xi}+\frac{1}{2\xi})}
&0\\
0&\frac{sin\frac{\pi}{2\xi}~cos\frac{i\pi\b}{2\hbar\xi}}
{sin\pi(\frac{i\b}{2\hbar\xi}+\frac{1}{2\xi})}
&\frac{cos\frac{\pi}{2\xi}~sin\frac{i\pi\b}{2\hbar\xi}}
{sin\pi(\frac{i\b}{2\hbar\xi}+\frac{1}{2\xi})}
&0\\
-\frac{sin\frac{\pi}{2\xi}~sin\frac{i\pi\b}{2\hbar\xi}}
{cos\pi(\frac{i\b}{2\hbar\xi}+\frac{1}{2\xi})}
&0&0&
\frac{cos\frac{\pi}{2\xi}~cos\frac{i\pi\b}{2\hbar\xi}}
{cos\pi(\frac{i\b}{2\hbar\xi}+\frac{1}{2\xi})}
\end{array}
\right).\label{V8}
\eea

\subsection{The ordinary trigonometric limit of $sl(n)$ elliptic R-matrix}
The ordinary trigonometric limit of the R-matrix (\ref{ER}) is
taken by:  $\tau \longrightarrow +i\infty$ with $z$ and 
$w$ fixed \cite{Ric86,Shi92}. Noting the properties 
\bea
\frac{\t\left[
\begin{array}{c}
\frac{1}{2}+\frac{a}{n}\\
\frac{1}{2}
\end{array}
\right](z,\tau)}
{\t\left[
\begin{array}{c}
\frac{1}{2}+\frac{a}{n}\\
\frac{1}{2}
\end{array}
\right](w,\tau)}
~&\stackrel{\tau\rightarrow +i\infty}{\longrightarrow}&~
\left\{\begin{array}{lll}
exp\{2i\pi(\frac{a}{n}-\frac{1}{2})(z-w)\},&&0<a<n\\
exp\{2i\pi(\frac{a}{n}+\frac{1}{2})(z-w)\},&&-n<a<0
\end{array}
\right.,\no\\
\frac{\t\left[
\begin{array}{c}
\frac{1}{2}\\
\frac{1}{2}
\end{array}
\right](z,\tau)}
{\t\left[
\begin{array}{c}
\frac{1}{2}\\
\frac{1}{2}
\end{array}      
\right](w,\tau)}
~&\stackrel{\tau\rightarrow +i\infty}{\longrightarrow}&~
\frac{sin\pi~z}{sin\pi~w},\no
\eea
we have that in ordinary trigonometric limit the matrix elements of 
(\ref{ER}) become\cite{Shi92}
\bea
\{R^{Q}\}^{kl}_{ij}(\b, \xi;\hbar)&\stackrel{def}{=}&
\k(\b)\times\lim_{\tau\rightarrow
+i\infty}\S^{kl}_{ij}(\frac{i\b}{\hbar\xi},\frac{1}{\xi},\tau)
\label{tlim}\\
&=& \left\{
\begin{array}{ll}
\{R^{Q}\}^{ii}_{ii}(\b, \xi;\hbar)=\k(\b)&\\
\{R^{Q}\}^{ij}_{ij}(\b,\xi;\hbar)
=\k(\b)~\frac{sini\pi\b}{sin\pi(\frac{i\b}{\hbar\xi}+\frac{1}{\xi})}
exp\{2i\pi(\frac{j-i}{n}-\frac{1}{2})\frac{i\b}{\hbar\xi}\},&~~i<j
\\
\{R^{Q}\}^{ij}_{ij}(\b,\xi;\hbar)
=\k(\b)~\frac{sini\pi\b}{sin\pi(\frac{i\b}{\hbar\xi}+\frac{1}{\xi})}
exp\{2i\pi(\frac{j-i}{n}+\frac{1}{2})\frac{i\b}{\hbar\xi}\},&~~j<i
\\
\{R^{Q}\}^{ij}_{ji}(\b, \xi;\hbar)
=\k(\b)~\frac{sini\pi\b}{sin\pi(\frac{i\b}{\hbar\xi}+\frac{1}{\xi})}
exp\{2i\pi(-\frac{1}{2} +\frac{j-i}{n})\frac{i\b}{\hbar\xi}\},&~~i<j\\
\{R^{Q}\}^{ij}_{ji}(\b, \xi;\hbar)
=\k(\b)~\frac{sini\pi\b}{sin\pi(\frac{i\b}{\hbar\xi}+\frac{1}{\xi})}
exp\{2i\pi(\frac{1}{2} +\frac{j-i}{n})\frac{i\b}{\hbar\xi}\},&~~j<i
\\
\{R^{Q}\}^{kl}_{ij}(\b, \xi;\hbar)=0,&~~otherwise
\end{array}
\right.
.\no
\eea
In the particular case n=2, one recovers 
\bea 
R^{Q}(\b,\xi;\hbar)=\k(\b)\left(
\begin{array}{cccc}
1&0&0&0\\0&
\frac{sin\frac{i\pi\b}{\hbar\xi}}{sin\pi(\frac{i\b}{\hbar\xi}+\frac{1}{\xi})}
&\frac{sin\frac{i\pi}{\xi}}{sin\pi(\frac{i\b}{\hbar\xi}+\frac{1}{\xi})}
&0\\0&
\frac{sin\frac{i\pi}{\xi}}{sin\pi(\frac{i\b}{\hbar\xi}+\frac{1}{\xi})}&
\frac{sin\frac{i\pi\b}{\hbar\xi}}{sin\pi(\frac{i\b}{\hbar\xi}+\frac{1}{\xi})}
&0\\
0&0&0&1
\end{array}
\right).\label{V6}
\eea
In the case $\hbar=\pi$, the above R-matrix coincides with the S-matrix
of sine-Gordon model\cite{Sim92}.

\sect{Modular transformation and the twist from $R^{Q}$ to
$R_{\xi}^{DY}$}
\subsection{Modular transformation of elliptic functions}
The elliptic $\s_0$-function has the following modular transformation
property\cite{Mum83}
\bea
\t \left[\begin{array}{c}
\frac{1}{2}\\
\frac{1}{2}
\end{array}
\right]
(\frac{z}{\tau},-\frac{1}{\tau})=exp\{i\pi\frac{z^2}{\tau}\} 
\t \left[\begin{array}{c}
\frac{1}{2}\\              
\frac{1}{2}
\end{array}
\right](z,\tau)\times const.\no
\eea
where the $const.$ only depends on $\tau$. Noting that 
\bea
\t \left[
\begin{array}{l}
\frac{1}{2}+a\\\frac{1}{2}+b
\end{array}
\right](z,\tau)=exp\{2i\pi[z+\frac{1}{2}+b+\frac{a\tau}{2}]\}
\t \left[
\begin{array}{l}
\frac{1}{2}\\\frac{1}{2}
\end{array}
\right](z+b+a\tau,\tau),\no
\eea
we have the following properties under modular transformation
\bea
\t \left[\begin{array}{c}
\frac{1}{2}+a\\
\frac{1}{2}+b
\end{array}
\right]  
(\frac{z}{\tau},-\frac{1}{\tau})=exp\{i\pi[\frac{z^2}{\tau}+a-b+2ab]\}
\t \left[\begin{array}{c}   
\frac{1}{2}+b\\
\frac{1}{2}-a
\end{array}
\right](z,\tau)\times const..\label{MT1}
\eea
From (\ref{MT1}) and the definition of $\Z$-symmetric R-matrix, we can
derive the following relations\cite{Fan98}
\bea
(M\otimes M)\S(\frac{z}{\tau},\frac{w}{\tau},-\frac{1}{\tau})
(M^{-1}\otimes
M^{-1})=x^{\frac{2zw(1-n)}{n\tau}}P\S(z,w,\tau)P.\label{MT2}
\eea
where $P$ is the permutation operator acting on  tensor space
${\bf V}\otimes {\bf V}$ as: $P(e_i\otimes e_j)=e_j\otimes e_i$, and 
the $n\times n$ matrix $M$ with the elements defined as 
$(M)_{jk}=\o^{-jk}$. The matrix $M$ enjoys in the following properties 
\bea
MgM^{-1}=h^{-1},~~~ MhM^{-1}=g.
\eea
\subsection{Twist from $R^{Q}$ to $R_{\xi}^{DY}$}
In this subsection, we shall calculate the twist from $\DR$ defined in 
(\ref{slim}) to $R^{Q}$ given in (\ref{tlim}). From (\ref{slim}), we have
\bea
\DR(\b,\xi;\hbar)&=&\k(\b)\lim_{w\rightarrow 0+}\S(\frac{i\b
w}{\hbar},w,\xi w)\no\\
&=&\k(\b)(M\otimes
M)P\lim_{w\rightarrow
0+}\{\S(\frac{i\b}{\hbar\xi},\frac{1}{\xi},-\frac{1}{\xi
w})\}P(M^{-1}\otimes M^{-1})\no\\
&=&\k(\b)(M\otimes 
M)P\lim_{\tau\rightarrow 
+i\infty}\{\S(\frac{i\b}{\hbar\xi},\frac{1}{\xi},\tau)\}P(M^{-1}\otimes
M^{-1})\no\\ 
&=&(M\otimes M)PR^{Q}(\b,\xi;\hbar)P(M^{-1}\otimes M^{-1}).\no
\eea
In the second line we have used (\ref{MT2}). Therefore, the trigonometric
degeneracy limits R-matrices $R^{DY}_{\xi}$ and $R^{Q}$ are related by 
\bea 
R^{DY}_{\xi}(\b,\xi;\hbar)=F_{21}R^{Q}(\b,\xi;\hbar)F_{12}^{-1},
\eea
where the twist $F_{12}$ is 
\bea
F_{12}=M\otimes M P_{12}.
\eea
In the following, let us consider the special case $n=2$. In this case,
\bea 
M=\left(\begin{array}{cc}
-1&1\\1&1\end{array}
\right)=-\sqrt{2}V\s_z,\no
\eea
where 
\bea
\s_z=\left(\begin{array}{cc}1&\\&-1\end{array}\right),~~~
V=\frac{1}{\sqrt{2}}\left(\begin{array}{cc}1&1\\-1&1\end{array}\right).\no
\eea
Thanks to that $R^{Q}$ (\ref{V6}) satisfies the following properties
\bea
PR^Q(\b,\xi;\hbar)P=R^{Q}(\b,\xi;\hbar)
=(\s_z\otimes\s_z)R^{Q}(\b,\xi;\hbar)(\s_z\otimes\s_z),\no
\eea
our twist is equivalent to that given by Arnaudon et al\cite{Arn00}.

\vskip.3in
\subsection*{ Acknowledgments.}

We would like to thank Bo-yu Hou for his encouragement. 
W.-L. Yang would like to thank Prof. von Gehlen and the
theoretical group of the Physikalishes Institut der
Universit\"at Bonn for their kind hospitality. 
This work has been partly supported by the National Natural
Science  Foundation of China. W.-L. Yang is supported by the
Alexander von Humboldt Foundation.

%\newpage

\end{document}